\documentclass[conference]{IEEEtran}

\IEEEoverridecommandlockouts
\usepackage{cite}
\usepackage{amsmath,amssymb,amsfonts}
\usepackage{algorithmic}
\usepackage{graphicx}
\usepackage{textcomp}
\usepackage{xcolor}
\usepackage{algorithmic}
\usepackage{graphicx}
\usepackage{subcaption}
\usepackage{textcomp}
\usepackage{todonotes}
\usepackage{xcolor}
\usepackage{hyperref}
\usepackage{cleveref}
\usepackage[most]{tcolorbox}
\usepackage[inline]{enumitem}
\usepackage{multirow}
\usepackage{algorithmic}
\usepackage{stfloats}
\usepackage[font=normalsize]{caption}
\usepackage[font=bf]{caption}
\usepackage{booktabs}
\AtBeginDocument{%
  \providecommand\BibTeX{{%
  Bib\TeX}}}

\newcounter{mainfindingid}
\newcommand{\defmainfinding}[3]{\refstepcounter{mainfindingid}\label{#1}\textbf{Main Finding \arabic{mainfindingid}:} #2

\textbf{Actionable Insight \arabic{mainfindingid}:} #3}

\newcounter{observationid}
\newcommand{\defobservation}[3]
{\refstepcounter{observationid}\label{#1}\textbf{Observation 
\arabic{observationid}:} #2}

\usepackage{amssymb}%
\usepackage{pifont}%
\newcommand{\cmark}{\ding{51}}%
\newcommand{\xmark}{\ding{55}}%

\usepackage{xcolor}
\usepackage{hyperref}

\begin{document}

\def\BibTeX{{\rm B\kern-.05em{\sc i\kern-.025em b}\kern-.08em
    T\kern-.1667em\lower.7ex\hbox{E}\kern-.125emX}}

\title{Generic and ML Workloads in an HPC Datacenter: Node Energy, Job Failures, and Node-Job Analysis}

\author{%
\IEEEauthorblockN{
Xiaoyu Chu$^{*1}$, Daniel Hofstätter$^{*2}$, Shashikant Ilager$^{2}$, Sacheendra Talluri$^{1}$, \\
Duncan Kampert$^{3}$, Damian Podareanu$^{3}$, Dmitry Duplyakin$^{4}$, Ivona Brandic$^{2}$, Alexandru Iosup$^{1}$}%
\IEEEauthorblockA{$^{1}$Vrije Universiteit, Amsterdam, the Netherlands \\
$^{2}$TU Wien, Vienna, Austria \\
$^{3}$SURF, Amsterdam, the Netherlands \\
$^{4}$National Renewable Energy Laboratory, Colorado, USA \\
\{x.chu, s.talluri, a.iosup\}@vu.nl, \{daniel.hofstaetter, shashikant.ilager, ivona.brandic\}@tuwien.ac.at, \\
\{duncan.kampert, damian.podareanu\}@surf.nl, \{dmitry.duplyakin\}@nrel.gov}

\thanks{*Equal contributions, joint first authors.}%
}

\maketitle
\thispagestyle{plain}
\pagestyle{plain}

\begin{abstract}
HPC datacenters offer a backbone to the modern digital society. 
Increasingly, they run Machine Learning~(ML) jobs next to generic, compute-intensive workloads, supporting science, business, and other decision-making processes.
However, understanding how ML jobs impact the operation of HPC datacenters, relative to generic jobs, remains desirable but understudied.
In this work, we leverage long-term operational data, collected from a national-scale production HPC datacenter, and statistically compare how ML and generic jobs can impact the performance, failures, resource utilization, and energy consumption of HPC datacenters.
Our study provides key insights, e.g., 
ML-related power usage causes GPU nodes to run into temperature limitations, 
median/mean runtime and failure rates are higher for ML jobs than for generic jobs,
both ML and generic jobs exhibit highly variable arrival processes and resource demands, significant amounts of energy are spent on unsuccessfully terminating jobs,
and concurrent jobs tend to terminate in the same state.
We open-source our cleaned-up data traces on Zenodo (\url{https://doi.org/10.5281/zenodo.13685426}), and provide our analysis toolkit as software hosted on GitHub (\url{https://github.com/atlarge-research/2024-icpads-hpc-workload-characterization}).
This study offers multiple benefits for data center administrators, who can improve operational efficiency, and for researchers, who can further improve system designs, scheduling techniques, etc.

\end{abstract}


\section{introduction}\label{sec1:introduction}\label{sec:introduction}

High Performance Computing (HPC) datacenters are important Information and Communications Technology (ICT) infrastructures for our society, particularly for scientific research and its many applications. Contemporary HPC datacenters are well-designed and highly tuned for reliable and resource-efficient execution of CPU-based scientific computing workloads~\cite{DBLP:conf/sc/GuptaPET17,DBLP:journals/tdsc/SchroederG10, li2022ai, patel2019perq}. %
However, as HPC datacenters are increasingly hosting Machine Learning (ML) jobs, understanding the different requirements and usage characteristics of such jobs is important for the design and tuning of future HPC datacenters. Early studies identify new energy~\cite{power_jobs_ipdps_20} and failure~\cite{DBLP:conf/sc/ShinOKEW21} patterns; existing schedulers and workload managers (e.g., SLURM) do not take into account the unique needs of these new types of jobs~\cite{li2022ai, weng2022mlaas}, potentially leading to substantial waste of researchers’ time, and computing energy resources\cite{DBLP:conf/usenix/AmvrosiadisPGGB18, ilager_dc_analysis_UCC23, zeus_nsdi_23}. 
Addressing the challenges of comparing ML and generic workloads in HPC datacenters, in this study,  we thoroughly investigate node energy, job failures, and joint node-job analysis. In this process, addressing the further challenge of data scarcity from HPC datacenters hosting both ML and generic HPC workloads, we collect and open-source long-term job and node data from a production, national-scale HPC datacenter. Our work leads to various findings and actionable insights, and eventually to stronger capabilities to improve resource allocation policies and job management strategies for combined ML and generic HPC workloads~\cite{DBLP:conf/osdi/QiaoCSNH0GX21, DBLP:conf/osdi/MohanPKC22}.

\begin{figure}[!t]
  \centering
  \vspace*{-0.0cm}
  \includegraphics[width=\linewidth]{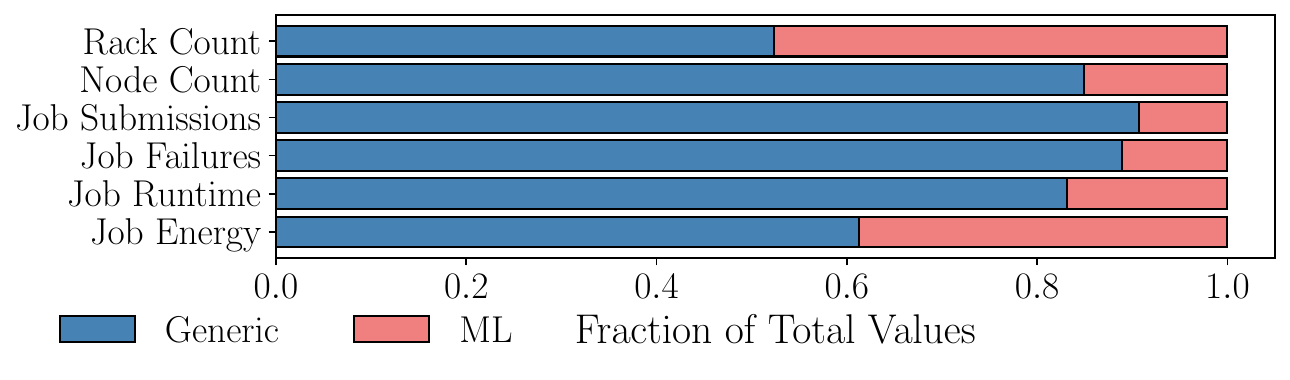}
  \vspace*{-0.5cm}
  \caption{Generic vs. ML hardware and workload, summary. Energy demands of ML jobs are proportionally higher than their share of submissions and runtime.}
  \label{fig:page1-comparison}
  \vspace*{-0.5cm}
\end{figure}

\textbf{It is important to systematically characterize ML and generic workloads running in HPC systems, to understand resource usage behavior, failure patterns, distributions, and correlations across different dependent parameters.}
Previous studies on HPC trace analysis have primarily focused on job-level data \cite{DBLP:conf/hipc/PaulFMGMB20, DBLP:conf/cluster/CarnsLRILR09, DBLP:conf/sc/PatelLKRAT20} or machine-level data \cite{tiwari2015understanding}, without considering the interplay between them. However, this can lead to inaccurate insights, as the performance of HPC jobs is highly influenced by the machine and infrastructure conditions \cite{DBLP:conf/dsn/El-SayedS13, DBLP:journals/ijhpca/Snir, white2017challenges, DBLP:conf/hipc/PaulFMGMB20}. This can be evidenced by the exemplary results from our analysis in \Cref{fig:page1-comparison}, which shows ML jobs running on GPU-accelerated nodes consume 39\% of the datacenter's total energy, even though they make up only 15\% of the datacenter's nodes and only about 9\% of the workload submissions. In addition, ML workloads experience slightly more failures %
than their node-count indicates.

\begin{table*}[t]
    \renewcommand{\arraystretch}{1.3}
    \caption{Cluster overview, shows the total size of the cluster in our study and typical per-node configurations.}
    \label{table:cluster-overview}
    \vspace{-0.2cm}
    \centering
    \resizebox{\textwidth}{!}{
    \begin{tabular}{@{}rr|rrrrrrrr@{}}
    \toprule
    \textbf{}                     & \textbf{\#Nodes}  & \textbf{\#CPUs} & \textbf{\#CPU Cores} & \textbf{CPU TDP} & \textbf{Memory} & \textbf{Storage} & \textbf{\#GPUs} & \textbf{GPU TDP} & \textbf{GPU Memory} \\ \midrule
    CPU-only Nodes               & 287              & 1               & 16                   & 125 W            & 96 GB           & 1.7 TB          & n/a               & n/a                & n/a                  \\
    GPU Nodes                    & 51               & 2               & 24                   & 210 W            & 192 GB          & 2.4 TB          & 4               & 1,120 W          & 96 GB               \\ \midrule
     Total  Values               & 338              & 489             & 5872                 & 51,425 W         & 46,336 GB       & 644.1 TB         & 198             & 53,040 W         & 3,712 GB            \\ \bottomrule
    \end{tabular}}
    \vspace{-0.25cm}
\end{table*}

\textbf{The objective of this paper is to present an in-depth HPC datacenter analysis and study the characteristics of generic and ML jobs using operational logs.} 
We collected and open-sourced detailed, long-term, job and node-level monitoring information from a production, national-level HPC datacenter---in total, approximately 94 million tuples with 100 metrics, covering four overlapped months of job and node data. 
We first performed data cleaning and preprocessing (integration) steps, followed by a detailed statistical analysis. We used various methods of data analysis, such as basic statistics, temporal patterns (e.g., trend analysis), distributions (e.g., Probability Density Function (PDF), or Cumulative Distribution 
Function (CDF)), and Pearson correlations. %
Our key contributions are: %
\begin{enumerate}
     
    \item %

    We propose a data processing and characterization method (\Cref{sec:method}) for comparing generic and ML workloads, offering insights into large-scale infrastructure by integrating long-term, high-quality node and job data.
  
    \item 
    We unveil how hardware and workloads differ in a heterogeneous HPC environment. Therefore, we study the cluster hardware utilization (\Cref{sec4:analysis-of-resource-usage}), analyze the characteristics and failure patterns of generic and ML jobs (Section \ref{sec5:jobfailures}), and investigate energy usage and correlations among generic and ML job types~(Section \ref{sec:cross-analysis}).

    \item We contribute to open-science by publishing job and node monitoring data from a relevant HPC datacenter (\url{https://doi.org/10.5281/zenodo.13685426})~\cite{chu_2024_13685426} and the analysis software toolkit (\url{https://github.com/atlarge-research/2024-icpads-hpc-workload-characterization}), ensuring reproducibility and supporting further research.

\end{enumerate}

\begin{table}[t] %
    \caption{Rack-level overview, shows the size and typical per-rack configurations.}
    \label{table:rack-overview}
    \vspace{-0.2cm}
    \centering
    \resizebox{\linewidth}{!}{
    \begin{tabular}{@{}rr|rrrrrr@{}}
    \toprule
    \textbf{}                 & \textbf{\#Racks} & \textbf{\#Nodes} & \textbf{CPU TDP}  & \textbf{GPU TDP}    \\ \midrule
    CPU-only Racks & 11    & 32              & 4,000 W           & n/a                   \\
    GPU Racks & 10        & 5               & 1,050 W           & 5,600 W             \\ \bottomrule
    \end{tabular}}
    \vspace{-0.25cm}
\end{table}

\section{System Background}\label{sec2:system-backgroud}

SURF Lisa is a Dutch national-scale datacenter consisting of 338 nodes distributed across 21 racks. Universities and researchers use it for different jobs, including bags of tasks, workflows, and ML training jobs. The jobs are submitted to a SLURM scheduler which then schedules them onto the nodes of the HPC cluster. A job can use a single node or multiple nodes. 
GPU nodes handle ML workloads for the vast majority of jobs (over 90\%), as indicated by the libraries (e.g., \textit{torch}, \textit{cuda}) used by each job, identified via \textit{XALT} by system administrators.
Our nodes employ various \textit{Second Generation Intel Xeon} processor models, \textit{NVIDIA TITAN RTX} or \textit{NVIDIA GeForce GTX 1080 Ti} GPUs, all installed in \textit{Dell EMC PowerEdge T640} node enclosures. Typical node configurations can be found in Table~\ref{table:cluster-overview}.

In our HPC datacenter, each rack comprises multiple nodes of the same type, i.e., CPU-only racks and GPU racks. A rack can host up to 32 CPU-only (generic) nodes or 7 GPU (ML) nodes, with the most common configurations listed in Table~\ref{table:rack-overview}. 
The rack air cooling is designed for a 5,500 W capacity. Noteworthy, while CPU-only racks remain within this cooling limit (CPU TDP), GPU racks (CPU+GPU TDPs) often exceed it due to the GPUs' high power demands.

\section{Characterization Method}\label{sec3:method}\label{sec:method}

\begin{table*}[!t] %
    \renewcommand{\arraystretch}{1.3}
    \caption{Data preparation overview. This work uses 3 different datasets. Legend: \#M=Number of metrics, \#R=Number of rows in millions, \#S=Size of dataset.}
    \label{table:data-preparation}
    \vspace{-0.2cm}
    \centering
    \resizebox{\textwidth}{!}{
    \begin{tabular}{@{}crrrrrrrrrrr@{}}
    
        \toprule
        \multicolumn{1}{c}{\textbf{ID}} & \multicolumn{1}{c}{\textbf{Name}} & \multicolumn{1}{c}{\textbf{Source}} & \multicolumn{1}{c}{\textbf{Start}} & \multicolumn{1}{c}{\textbf{End}} & \multicolumn{1}{c}{\textbf{\#M}} & \multicolumn{1}{c}{\textbf{\#R}} & \multicolumn{1}{c}{\textbf{\#S}} & \multicolumn{1}{c}{\textbf{Description}} \\
        \midrule
        (a) & Job Dataset & SLURM & 2021-12-26 & 2022-11-01 & 9 & %
        1.60 M & 26 MB & ID, dates, node types, \#nodes, \#cores, state \\
        (b) & Node Dataset & Prometheus & 2022-06-30 &  2022-11-22 &  91 & %
        127.83 M & 16 GB & Node memory, network, power usage, etc. \\
        (c) & Joint Dataset & (a) join (b) & 2022-06-30 &  2022-11-01 & 100 & %
        93.95 M & 10 GB & Information per-node and related jobs\\
        \bottomrule
    \end{tabular}}
\end{table*}

We propose a data-driven characterization method for analyzing and comparing generic and ML workloads, built mainly upon (1) hardware utilization and energy usage, (2) job failures and resource allocation patterns, and (3) joint analysis of node and job metrics. 
We also take a novel approach by correlating job exit states among concurrently running jobs.

\subsection{Data Collection} \label{sec:method:collection} %

We collected approximately 10 months of job data from SLURM, spanning from the end of December 2021 to November 2022.
Each job data point is sampled upon its termination, with information on resource allocation, runtime, and exit state. Additionally, we collected roughly 5 months of node data from Prometheus, ranging from June 2022 to November 2022. The sample interval for node data is set at 30s. The node dataset encompasses various software metrics such as packets received and I/O requests, alongside hardware metrics like CPU/GPU power and temperature. The low sampling interval and the large number of metrics presented in this dataset offer the potential for more in-depth and innovative insights into the operations of datacenters~\cite{DBLP:journals/fgcs/VersluisCGLPCUI23,li2022ai}, which we explore in this study. 

\begin{figure}[t]
    \centering
    \vspace*{-0.0cm}
    \includegraphics[width=\linewidth]{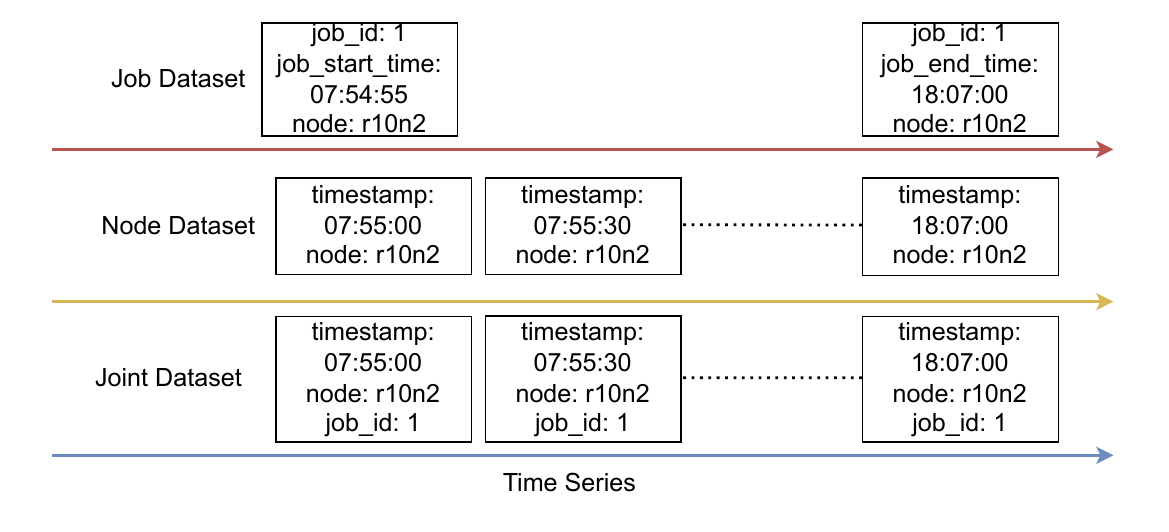}
    \vspace*{-0.5cm}
    \caption{An example of the data integration process. We match each job record to the fine granular 30s-interval timestamps of the node dataset.}
    \label{fig:data-integration}
\end{figure}

\subsection{Processing Metrics and Integrating Datasets} \label{sec:method:integration}

We aggregated raw JSON entries in each row by taking minimum, maximum, mean, and sum values to generate new attributes, e.g., the sum of all GPU power measurements per node and timestamp. 
Our cleaned node dataset has roughly 128 million tuples in total, with each having 91 features in it.

We integrated job data (Table \ref{table:data-preparation}-(a)) and node data (Table \ref{table:data-preparation}-(b)) to enable correlation and energy analysis across both levels. 
Therefore, we matched each job to corresponding node data logs over the job's duration. A similar approach is also taken in~\cite{DBLP:conf/sc/AnticiABK23} to calculate job energy usage utilizing node power metrics.
\Cref{fig:data-integration} illustrates an example of the data integration process. First, we take a job, e.g., job (job\_id: 1) which is executing on r10n2 (node) from 07:54:55 (job\_start\_time) to 18:07:00 (job\_end\_time). Second, for all node data samples within this period, the metrics for the job (job\_id: 1) are combined with the node data metrics. Afterward, we obtained the combined dataset (Table \ref{table:data-preparation}-(c)).
This method naturally leaves out jobs under 30 seconds that fall between node timestamps, but since they account for less than 0.1\% of the total runtime, their impact on energy consumption is negligible, which we mainly discuss by utilizing this combined dataset.

\section{Analysis of Node Utilization and Energy Usage}\label{sec4:analysis-of-resource-usage}\label{sec:analysis-of-power-usage}

In this section, we begin by analyzing patterns in overall cluster utilization using the node dataset (Table\ref{table:data-preparation}-(b)).

\begin{tcolorbox}[enhanced,breakable]
\defmainfinding{mf:resource-and-energy-usage}
{GPU nodes under-utilize their CPU (median of 9.6\% in \textit{Node Load 1} metric). Both CPU and GPU memory are rarely fully utilized. GPU temperature limits are reached regularly, and their temperatures are affected by hardware topology.}
{The CPU-GPU imbalance suggests that operators should provision imbalanced nodes with mixed jobs and save costs by adapting CPU configurations for GPU-heavy workloads. GPU performance can be improved by prioritizing the resource allocation of GPUs at positions with better cooling.}
\end{tcolorbox}

\begin{figure}[hbt!]
    \centering
    \vspace*{-0.0cm}
    \includegraphics[width=\linewidth]{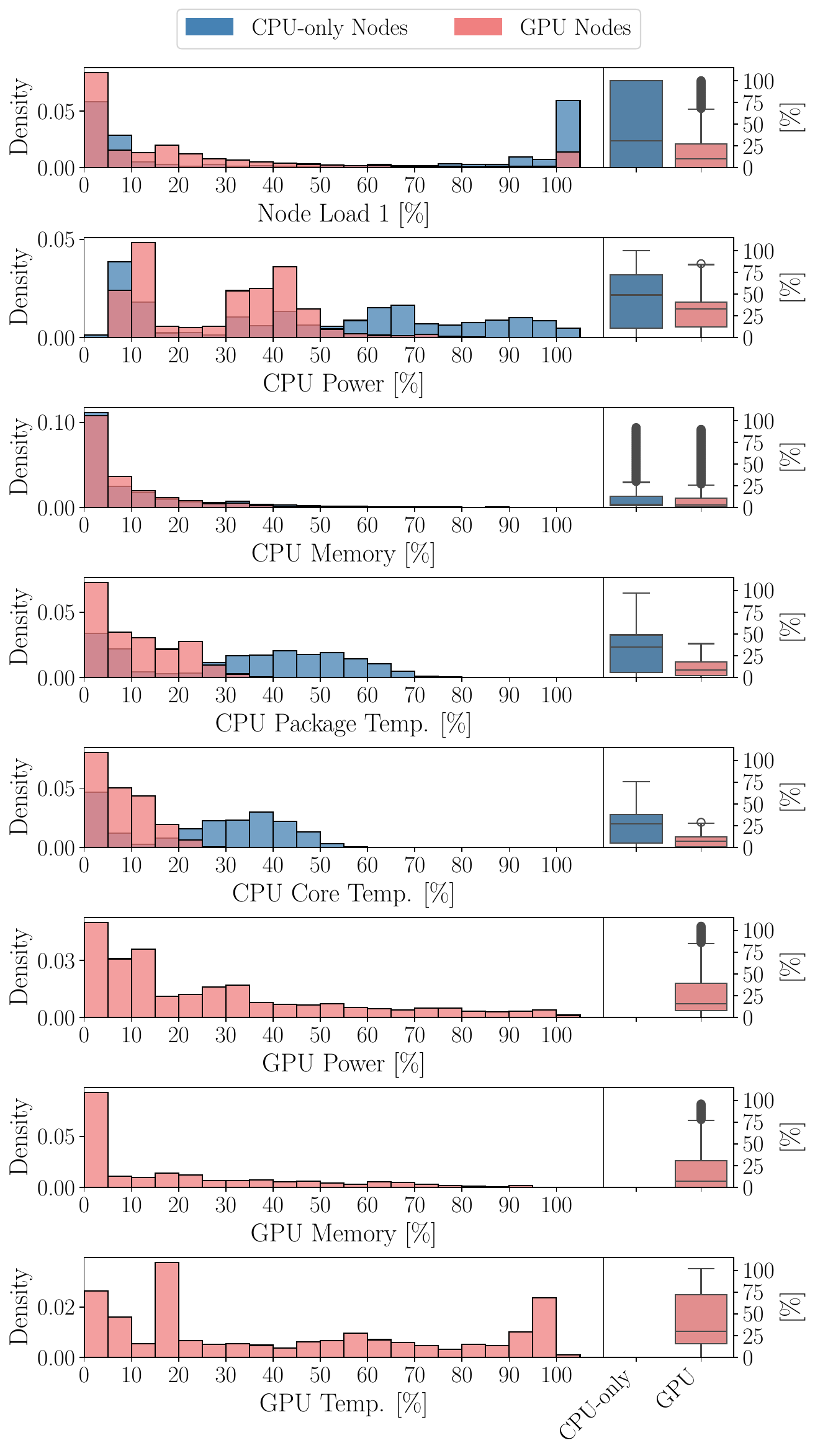}
  \caption{Normalized node utilization across various metrics is depicted using probability density functions (left) and box plots (right), revealing high GPU temperatures.}
  \label{fig:utilization_regular_vs_ml_distributions}
\end{figure}

\subsection{Node Utilization}

Figure~\ref{fig:utilization_regular_vs_ml_distributions} shows the probability distribution of various node attributes, normalized to a utilization value. Here, 100\% utilization is the maximum value according to hardware specifications, e.g., the TDP and memory configurations from Table~\ref{table:cluster-overview}, and the maximum allowed temperatures according to the CPU and GPU manufacturers. The \textit{Node Load 1} metric reflects the rolling average CPU thread load over the last minute.
We consider values equal to or larger than the CPU core count as 100\% utilization. We further clip all utilization values to their 99.99th percentile values, to reduce the impact of a few extreme outliers.

\vspace*{0.06cm}
\noindent\fbox{%
    \parbox{\linewidth}{%
    \defobservation{ob:}
    {CPUs and GPUs under-utilize memory, with mean  CPU memory usage below 11\% in both CPU-only and GPU nodes and mean VRAM (for GPUs) usage below 19\%.}
    }}
\vspace*{0.03cm}

CPU-only nodes show a higher thread load than GPU nodes (see Figure \ref{fig:utilization_regular_vs_ml_distributions}). 
CPU memory (RAM) is mostly under-utilized, with means of 10.3\% and 8.1\% for CPU-only and GPU nodes, respectively. While RAM is often under-utilized in datacenters~\cite{DBLP:journals/fgcs/VersluisCGLPCUI23,ilager_dc_analysis_UCC23,DBLP:conf/sbac-pad/PengPG20,li2023_nersc/10.1007/978-3-031-32041-5_16}, it may also be a bottleneck in other cases~\cite{DBLP:conf/iwqos/GuoCWDFMB19}. The mean GPU memory (VRAM) utilization of 18.7\% is higher, but still considerably low. However, low memory utilization may alone not be a sufficient metric for making a statement on memory over-provisioning, since peak loads also have to be handled~\cite{DBLP:conf/cluster/VermaKW14}, as evidenced by the many outliers towards 100\% memory utilization in the box plots of Figure~\ref{fig:utilization_regular_vs_ml_distributions}.

\vspace*{0.06cm}
\noindent\fbox{%
    \parbox{\linewidth}{%
    \defobservation{ob:}
    {GPUs reach temperature limitations regularly. 17\% of the time GPU temperature utilization exceeds 90\%.}
    }}
\vspace*{0.03cm}

Regarding CPU energy consumption, CPU-only nodes utilize the CPU more often at 100\% power than GPU nodes. Consequently, CPU package temperatures are also overall higher for CPU-only nodes. Individual CPU core temperatures show the same pattern and mostly follow the behavior of the CPU package. Due to higher allowed temperatures for individual cores of 101 °C, the lower threshold for package temperatures of 77-87 °C is reached earlier in most cases, meaning throttling of individual cores due to local hot spots is not an issue here.

Due to high GPU power utilization and over-provisioning of GPU TDPs in most GPU racks (Table \ref{table:rack-overview}), the limited cooling cannot keep up. As a result, GPU temperature utilization regularly reaches 100\% and is over 90\% about 17.4\% of the time. This raises the concern of thermal throttling, which limits GPU performance. One way of dealing with this issue is power-capping GPUs~\cite{li2022ai}, which not only helps to control power surges and temperatures but also reduces energy bills.

\subsection{Temperature Behavior of GPUs}

\begin{figure}[t!]
    \centering
    \includegraphics[width=\linewidth]{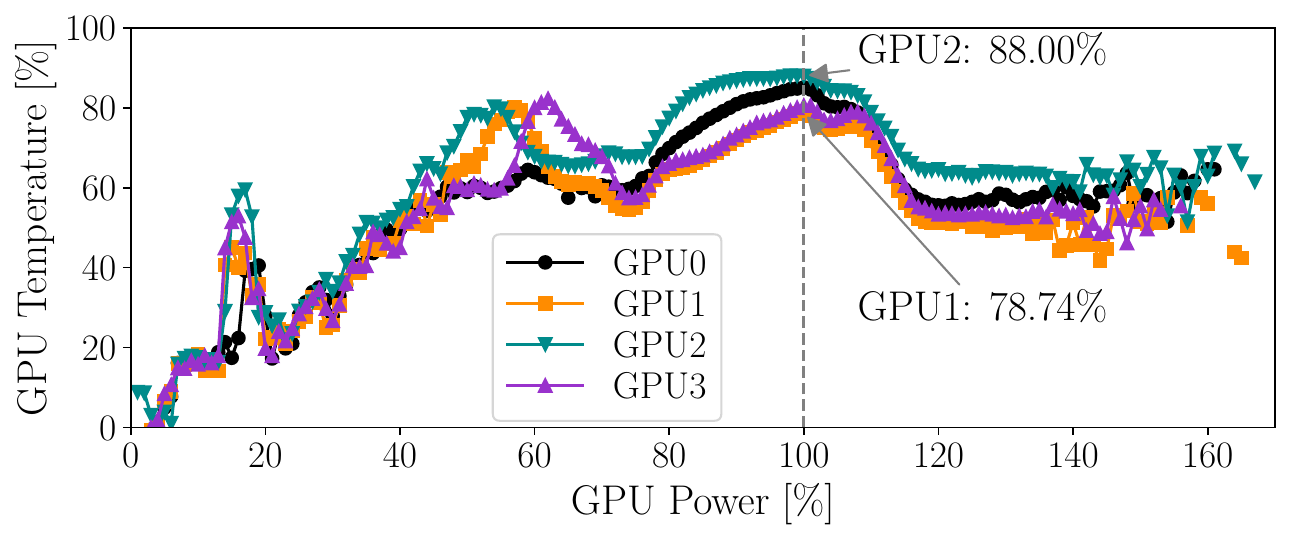}
  \caption{Average GPU temperature at various power utilizations across GPU indices (0 to 3) in the node. For the same power usage, GPU temperatures vary greatly.}
  \label{fig:gpu_power_temp}
\end{figure}

\begin{table}[t] %
\centering
\caption{Mean GPU power utilization.}
\vspace*{-0.2cm}
\label{tab:gpu_power}
\begin{tabular}{@{}lllll@{}}
\toprule %
\begin{tabular}[c]{@{}l@{}}GPU Name\end{tabular} & GPU0 & GPU1 & GPU2 & GPU3 \\ \hline
Mean Power Utilization                                          & 
34.27\%   & 25.06\%   & 28.02\% & 23.43\% \\ %
\bottomrule %
\end{tabular}
\end{table}

\vspace*{0.06cm}
\noindent\fbox{%
    \parbox{\linewidth}{%
    \defobservation{ob:}
    {GPU temperatures can vary significantly depending on their position inside the node, with differences of around 9\% in temperatures at 100\% power utilization.}
    }}
\vspace*{0.03cm}

Looking further into the temperature limitations of GPUs, Figure~\ref{fig:gpu_power_temp} shows how hardware topology affects GPU temperatures, aggregated over 48 GPU nodes equipped with 4 GPUs, where GPUs are indexed according to their physical arrangement inside the node. At 100\% power utilization (TDP of GPU), we observed significant differences in GPU temperatures, where GPU2 runs on average about 9\% hotter than GPU1, showing that the position of the GPU inside the node has a major influence on thermals. An explanation can be found by looking at the mechanical design of the used GPU models~\cite{titanrtx,1080ti} and node enclosures~\cite{dellt640}. 
Inside the node, GPUs are pairwise next to each other, potentially causing their fans to be partially obstructed and taking in hot air from the neighboring GPUs. This heat re-circulation effect has already been a discussed problem at the datacenter level~\cite{DBLP:journals/tpds/IlagerRB21}. Here, we evidenced similar phenomena in multi-GPU nodes.
Position-dependent thermal behavior of GPUs is also observed in \cite{DBLP:conf/sc/ShinOKEW21}, however, their study focused on water-cooled systems, while our work deals with air-cooled environments.

These insights indicate potential opportunities for optimizing GPU performance. Table~\ref{tab:gpu_power} reveals GPU0 is most utilized throughout the cluster. However, GPU1 and GPU3 would be better suited for higher workloads due to their superior cooling. Since GPUs can throttle due to temperature limitations, strategically assigning tasks to cooler GPUs could enhance performance. This can be achieved through, e.g., ML-based scheduling approaches that predict temperature~\cite{DBLP:journals/tpds/IlagerRB21}.

\section{Analysis of Job Characteristics and Failures}\label{sec5:jobfailures}
In this section, utilizing the job dataset (Table\ref{table:data-preparation}-(a)), we delve into the characterizations of job arrival times, wait times, run times, temporal patterns, and job sizes.

\begin{figure*}[!ht]
  \centering
  \begin{subfigure}[b]{0.49\textwidth}
    \includegraphics[width=\textwidth]{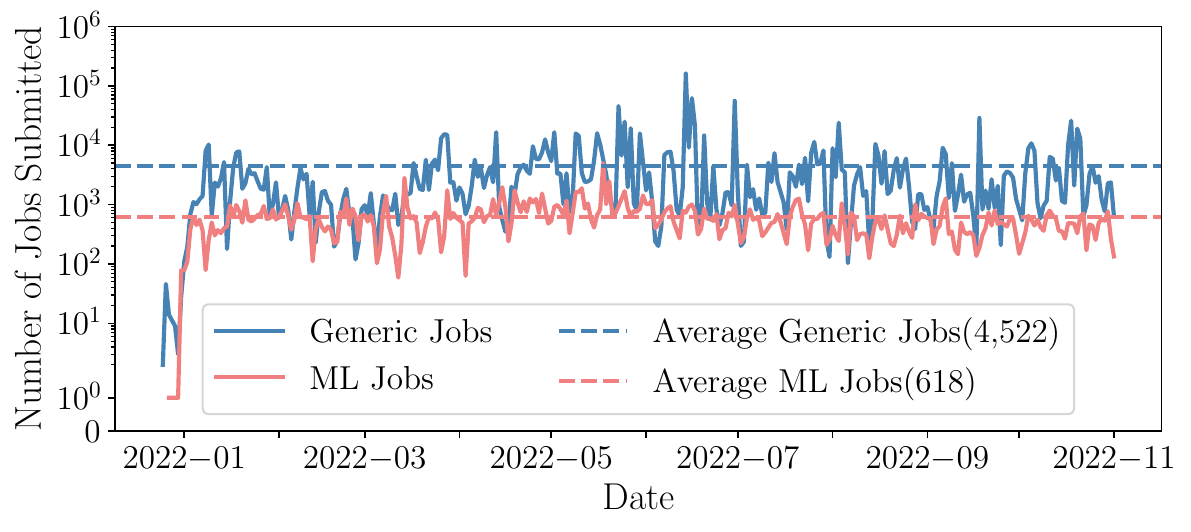}
    \caption{Jobs submitted by date.}
    \label{fig:job-by-date}
  \end{subfigure}
  \hfill
  \begin{subfigure}[b]{0.49\textwidth}
    \includegraphics[width=\textwidth]{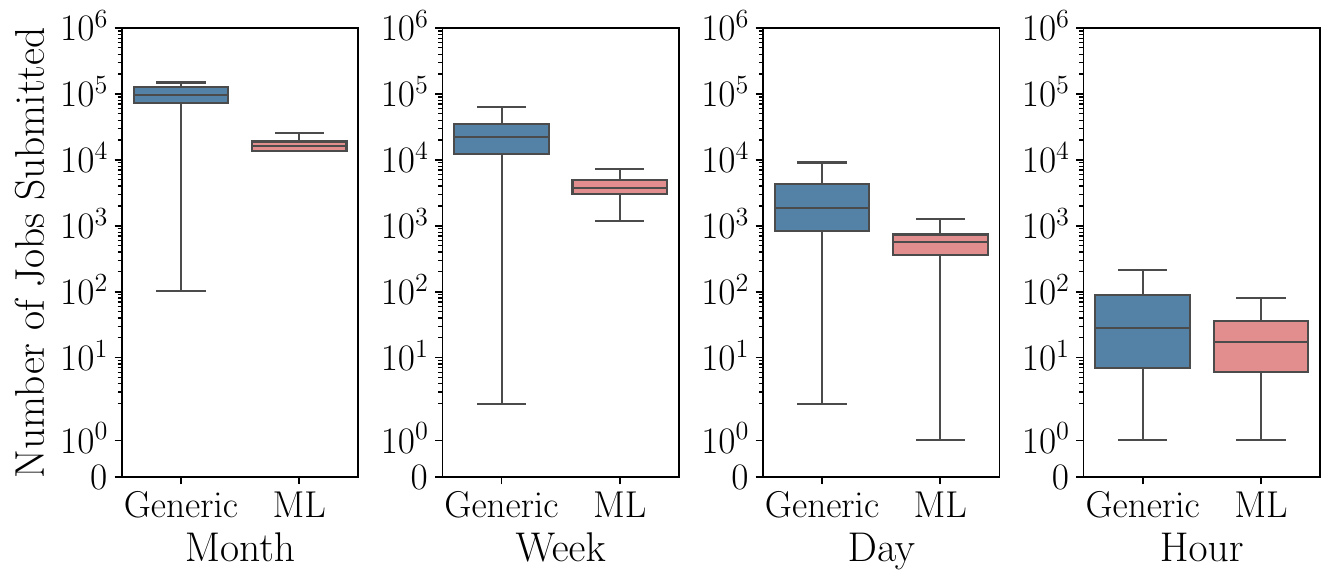}
    \caption{Jobs submissions, aggregated by different time granularity.}
    \label{fig:job-boxplot}
  \end{subfigure}
  \vspace{-0.15cm}
  \caption{The total number of submitted generic jobs and ML jobs, showing high variability over time.}
  \label{fig:number-of-submitted-job}
  \vspace{-0.15cm}
\end{figure*}

\begin{figure*}[t] %
  \centering
  \begin{subfigure}[b]{0.49\textwidth}
    \includegraphics[width=\textwidth]{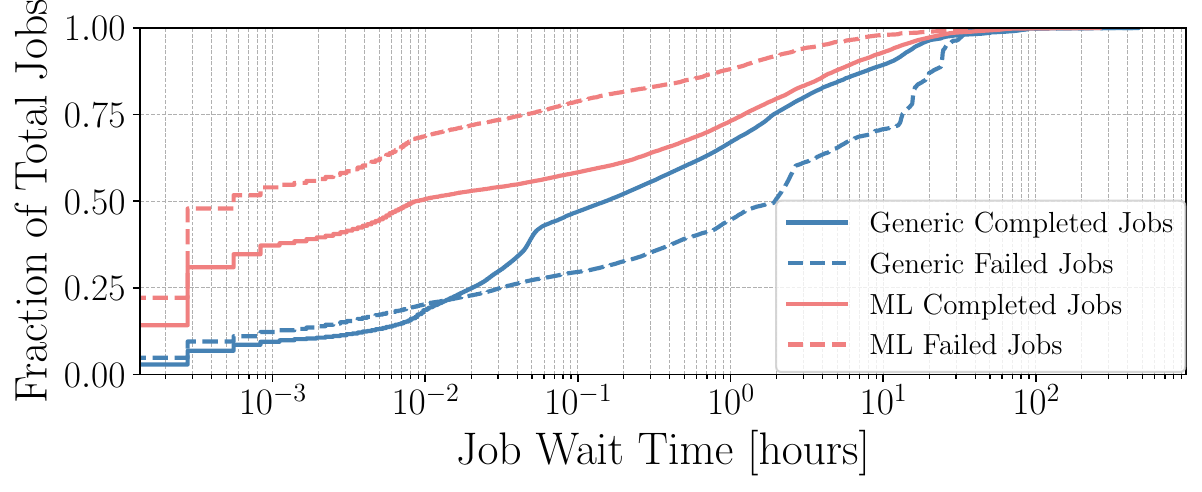}
    \caption{Job wait time, CDF plot.}
    \label{fig:job-wait-time}
  \end{subfigure}
  \hfill
  \begin{subfigure}[b]{0.49\textwidth}
    \includegraphics[width=\textwidth]{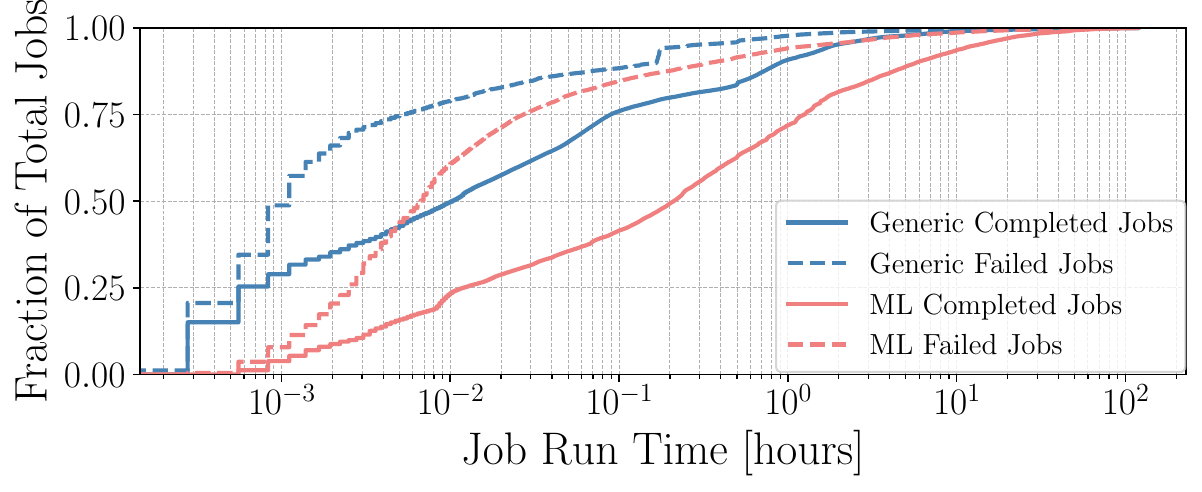}
    \caption{Job run time, CDF plot.}
    \label{fig:job-run-time}
  \end{subfigure}
  \vspace{-0.15cm}
  \caption{Job wait time and run time, showing ML jobs wait shorter but run longer than generic jobs.}
  \label{fig:job-time}
  \vspace*{-0.35cm}
\end{figure*}

\begin{tcolorbox}[enhanced, 
    breakable]
\defmainfinding{mf:performance}
{ML jobs have longer duration (median of 6.48 minutes) and smaller job sizes (average of 6.81 CPU cores) compared to generic jobs.}
{ML jobs' longer duration can be exploited to increase resource utilization with sophisticated scheduling  \& predictive dynamic optimizations (e.g., turning nodes on/off).  Smaller ML job sizes enable specialized network topologies \& placement algorithms that could avoid full bandwidth bisection~\cite{DBLP:conf/conext/ValadarskySDS16}.}
\end{tcolorbox}

\subsection{Job Arrivals}\label{sec:failures:job-arrival}
\vspace*{0.06cm}
\noindent\fbox{%
    \parbox{\linewidth}{%
    \defobservation{ob:}
    {Arrival and demand of generic and ML jobs are highly variable. The number of submitted jobs per day varies by up to three orders of magnitude for both. }
    }}
\vspace*{0.03cm}

Figure \ref{fig:job-by-date} gives a timeline overview of submitted generic and ML jobs based on dataset Table\ref{table:data-preparation}-(a). The number of submitted jobs per day is highly variable, as the curves are going up and down significantly between days for the 10 months.
On average, 4,522 generic jobs and 618 ML jobs are submitted daily.
Figure \ref{fig:job-boxplot} provides the distributions of job submissions across four types of time granularity, ranging from month to hour, excluding outliers based on the three-sigma ($3\sigma$) rule. While the median of hourly job submissions shows no significant difference, the maximum number of generic jobs exceeds that of ML jobs by a considerable margin. The daily count of both generic and ML jobs varies by up to three orders of magnitude. However, the number of job submissions is more steady for ML jobs compared to generic ones. The distribution can give insight into the datacenter simulator configuration~\cite{DBLP:conf/ccgrid/MastenbroekAJLB21}. Compared to the analysis results of the job data collected in 2020 from a similar study \cite{DBLP:journals/fgcs/VersluisCGLPCUI23}, the average amount of ML jobs increased from 320 to 618, reflecting the upward trend of ML research and application.

\subsection{Job Wait Time and Run Time}\label{sec:failures:wait-run-time}

\vspace*{0.06cm}
\noindent\fbox{%
    \parbox{\linewidth}{%
    \defobservation{ob:}
    {ML jobs have longer running times (2.71h) and shorter waiting times (1.84h) on average compared to generic jobs (0.83h and 4.21h, respectively).}
    }}
\vspace*{0.03cm}

We inspect the wait time and run time of generic and ML jobs, as shown in Figure \ref{fig:job-time}. Overall, the median waiting time for ML jobs (11.00 seconds) is less than for generic jobs (11.65 minutes). Additionally, completed jobs (7.93 minutes) have significantly shorter median waiting times than failed jobs (47.08 minutes). 
According to a study in \cite{DBLP:conf/sc/PatelLKRAT20} on the Argonne Leadership Computing Facility supercomputers, the median waiting time was approximately 1 hour in 2018, while in our cluster, it is 8.2 minutes.
The majority of generic jobs (64.67\%) and ML jobs (79.83\%) wait at most 1 hour.

Overall, the median running time for ML jobs (6.48 minutes) is about 16 times longer than for generic jobs (24 seconds).
In contrast, Li et al. \cite{li2023_nersc/10.1007/978-3-031-32041-5_16} report only a roughly 2-fold increase in median runtimes for GPU jobs compared to CPU jobs.
Additionally, in our cluster, completed jobs (44 seconds) have longer median running times than failed jobs (5 seconds). Around 85\% of ML jobs failed within 6 minutes and 94\% failed within 1 hour.
90.80\% of generic jobs are usually completed within 1 hour, whereas completed ML jobs have longer durations, with approximately 71.89\% taking at most 1 hour.
Amvrosiadis et al. \cite{DBLP:conf/usenix/AmvrosiadisPGGB18} report that 80\% of jobs have durations shorter than 12 minutes in their analysis of a Google trace, but significantly longer durations of 2-6 hours in 3 other traces they investigated. 
In contrast, the jobs in our cluster are shorter than 19 minutes at the same percentile, indicating that our job runtimes %
are between their extremes.

\subsection{Temporal Patterns of Job Failures}

\begin{figure*}[!t]
  \centering
  \begin{subfigure}[b]{0.49\textwidth}
    \includegraphics[width=\textwidth]{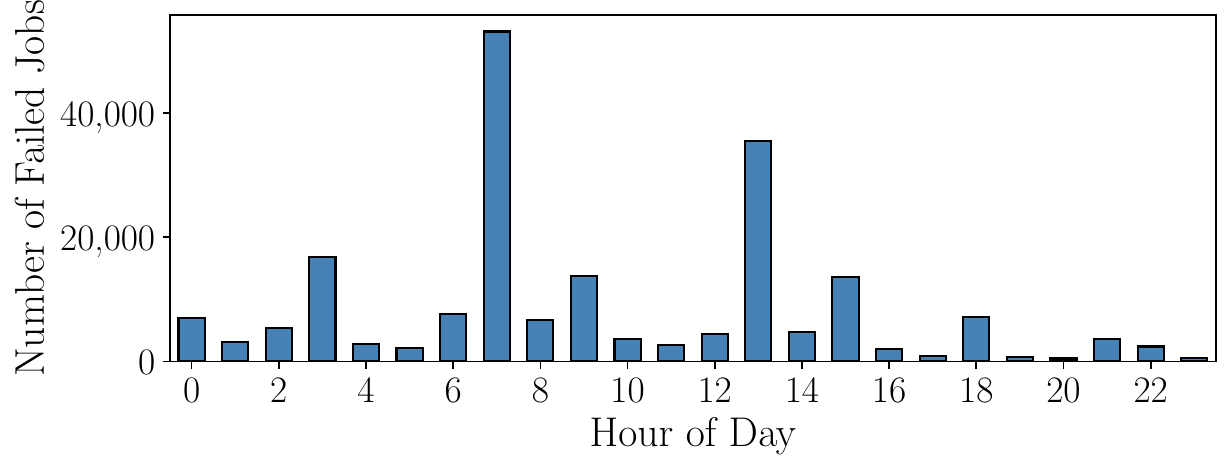}
    \caption{Generic job failures.}
    \label{fig:ge-failures-hour-of-day}
  \end{subfigure}
  \hfill
  \begin{subfigure}[b]{0.49\textwidth}
    \includegraphics[width=\textwidth]{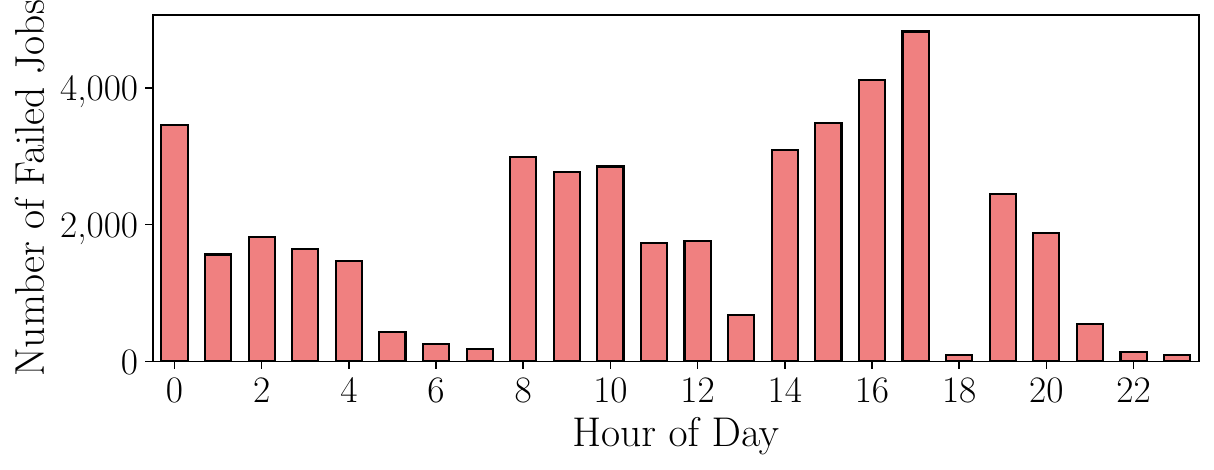}
    \caption{ML job failures.}
    \label{fig:ml-failures-hour-of-day}
  \end{subfigure}
  \vspace*{-0.15cm}
  \caption{The total number of failed jobs by hour of the day. Generic jobs show higher irregularities than ML jobs.}
  \label{fig:job-hour-of-day}
  \vspace*{-0.25cm}
\end{figure*}

\begin{figure*}[t] %
  \centering
  \begin{subfigure}[b]{0.32\textwidth}
    \includegraphics[width=0.98\linewidth]{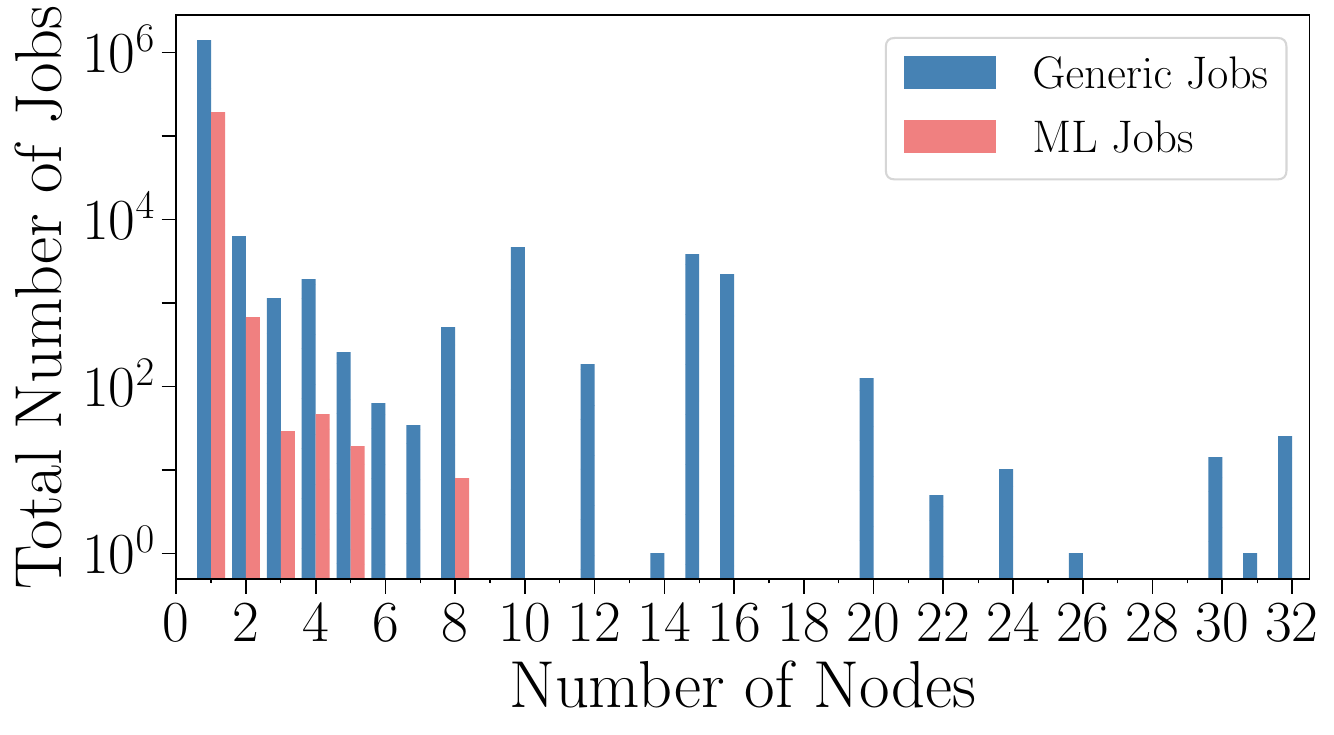}
    \caption{The number of nodes used per job.}
    \label{fig:job-nodes-usage}
  \end{subfigure}
  \hfill
  \begin{subfigure}[b]{0.32\textwidth}
    \centering
    \includegraphics[width=\linewidth]{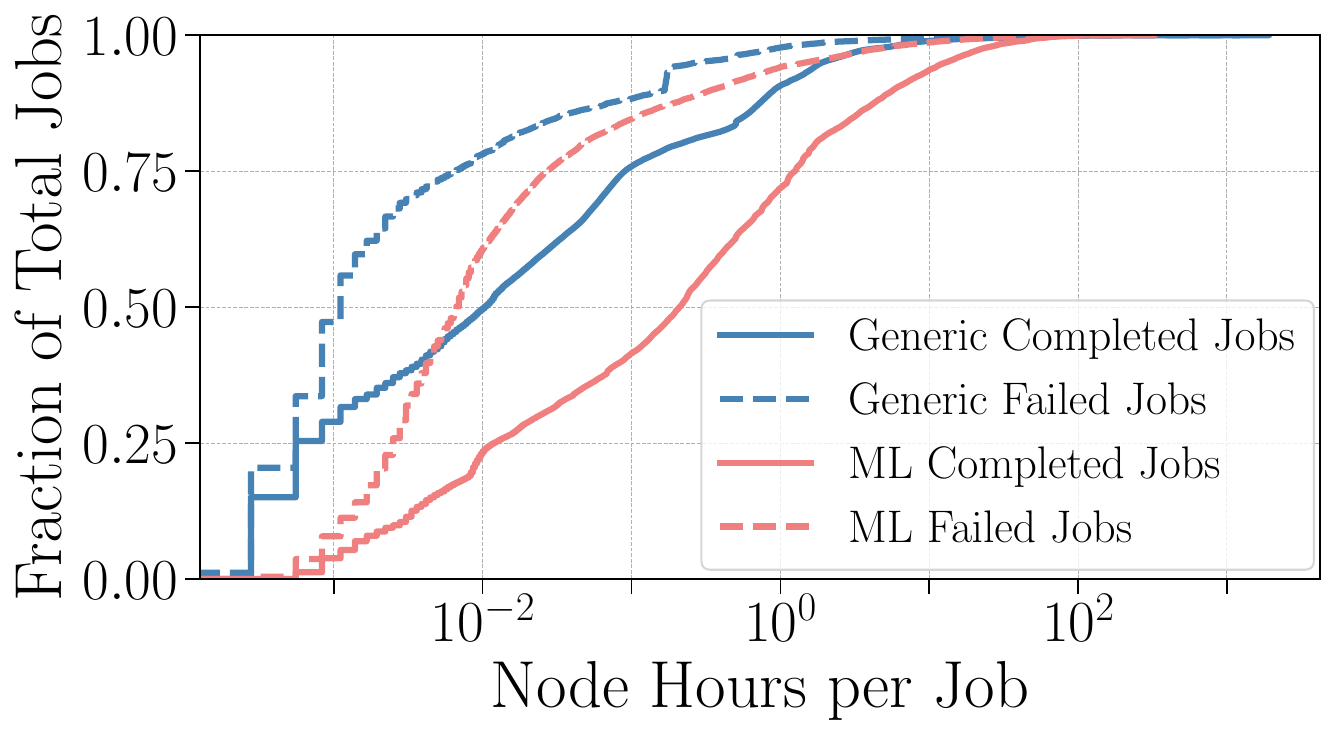}
    \caption{Node hours used per job.}
    \label{fig:job-node-hour}
  \end{subfigure}
  \hfill
  \begin{subfigure}[b]{0.32\textwidth}
    \includegraphics[width=\linewidth]{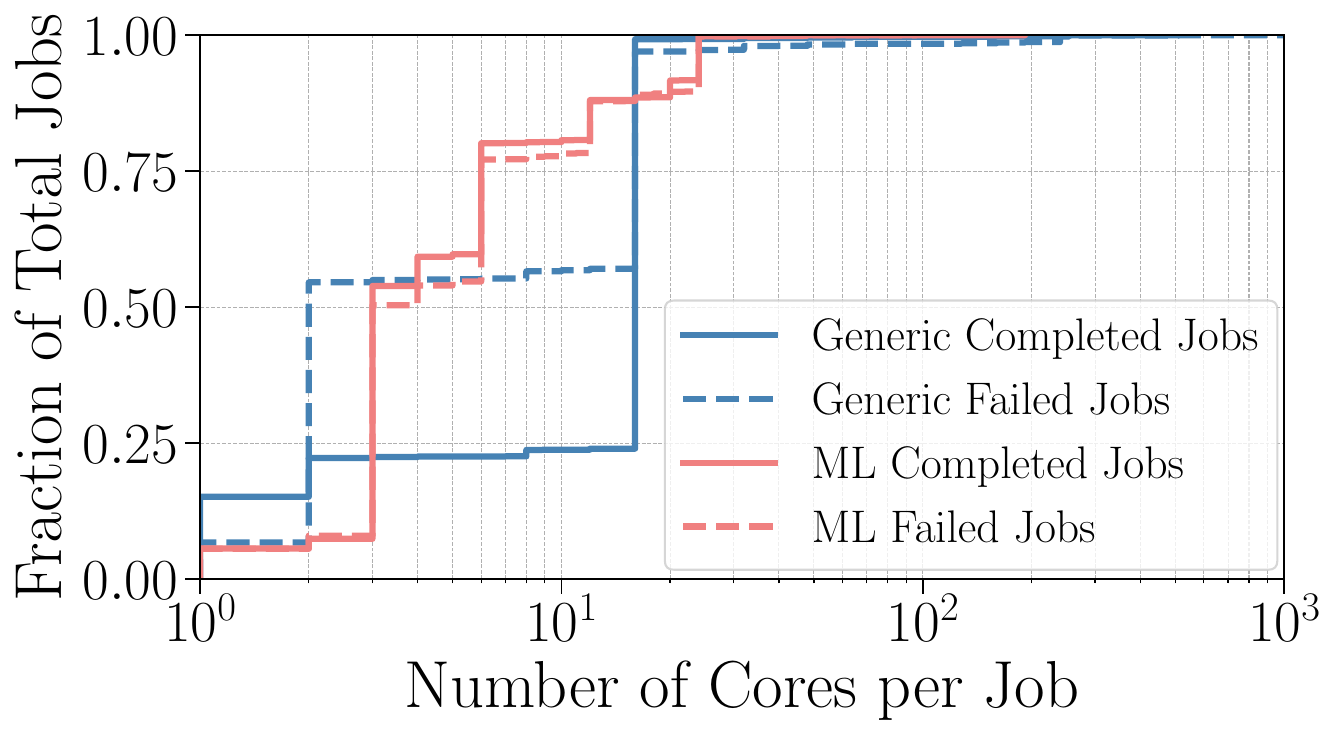}
    \caption{CPU cores used per job.}
    \label{fig:job-cores}
  \end{subfigure}
\vspace*{-0.15cm}
  \caption{Job size in nodes, node hours, and CPU cores. Generic jobs utilize more nodes and CPU cores than ML jobs.}
  \label{fig:job-size}
  \vspace*{-0.35cm}
\end{figure*}

\vspace*{0.06cm}
\noindent\fbox{%
    \parbox{\linewidth}{%
    \defobservation{ob:}
    {ML job failures have a diurnal pattern, whereas generic job failures exhibit irregular fail behavior, with anomaly peaks on certain days and hours.}
    }}
\vspace*{0.03cm}

To investigate the temporal patterns of generic and ML failed jobs, we aggregated job failures by the hour of the day, as shown in \Cref{fig:job-hour-of-day}.
Failures in ML jobs exhibit a daily pattern, typically occurring between 8 am and 5 pm. In contrast, generic job failures are irregular and erratic, with occasional spikes on specific days and hours. 
The daily patterns of ML job failures are similar to previous works \cite{DBLP:conf/wosp/ChuTVI23, DBLP:conf/usenix/AmvrosiadisPGGB18,DBLP:conf/sc/ShinOKEW21}. However, the pattern is inconsistent for generic job failures, which may be caused by our datacenter specific operational conditions.

\subsection{Job Size in Nodes, Node Hours, and Cores}\label{sec:failrues:job-size}

\vspace*{0.06cm}
\noindent\fbox{%
    \parbox{\linewidth}{%
    \defobservation{ob:}
    {ML jobs typically utilize fewer than 8 nodes, whereas generic jobs can utilize a variable number of nodes ranging from 1 to 32.}
    }}
\vspace*{0.03cm}

In \Cref{fig:job-size} we present the size of different jobs in terms of the number of nodes, node hours, and CPU cores.
\Cref{fig:job-nodes-usage} shows the distribution of node allocation per job. The majority of generic jobs (98.50\%) and ML jobs (99.60\%) only use a single node in the examined cluster. Despite that, generic jobs can utilize a variable number of nodes,
ranging up to 32,
while ML jobs utilize at most 8 nodes. For both types of jobs, the utilization of multiple nodes is around 1\%. 
Based on \Cref{fig:job-node-hour}, ML jobs use more median node hour time (6.5 minutes) than generic jobs (0.4 minutes), this is majorly due to the longer running time of ML jobs. Because most jobs only use one node, the squashed areas of node hours align with the result of job running time from \Cref{fig:job-run-time}.

\vspace*{0.06cm}
\noindent\fbox{%
    \parbox{\linewidth}{%
    \defobservation{ob:}
    {Completed generic jobs typically request more CPU core resources than failed ones, with both surpassing ML jobs in resource demand.}
    }}
\vspace*{0.03cm}

We also inspect the CPU cores used per job. \Cref{fig:job-cores} shows the CDF plot of user requests for CPU cores in successful and failed jobs. On average, generic jobs utilize more CPU cores (13.07) compared to ML jobs (6.81), which is expected since ML jobs mainly rely on GPUs for their computation. At the same time, there is no significant distinction between completed jobs (12.67) and failed jobs (10.99). The number of failed generic jobs sharply increases at 2 and 16 cores, indicating that 48\% and 40\% generic jobs failed at these core counts, respectively. A peak is also observed where 75\% completed generic jobs utilized 16 cores. We conjecture most users request one full node via the job scheduler (as indicated in \Cref{table:cluster-overview}, CPU-only nodes typically have 16 CPU cores). A similar pattern is also observed in earlier works~\cite{DBLP:conf/usenix/AmvrosiadisPGGB18, DBLP:journals/fgcs/VersluisCGLPCUI23}. However, failed and completed ML jobs have an unusual distribution of allocated CPUs, with nearly half (42\% and 46\% respectively) utilizing 3 cores. This is because the smaller ML jobs are provisioned 3 cores from a 24-core GPU node by the scheduler, confirmed by our datacenter operators.

\section{Joint Analysis of Job and Node Data}\label{sec:cross-analysis}

\begin{figure}[t] %
  \centering
  \begin{subfigure}[b]{\columnwidth} %
    \includegraphics[width=\textwidth]{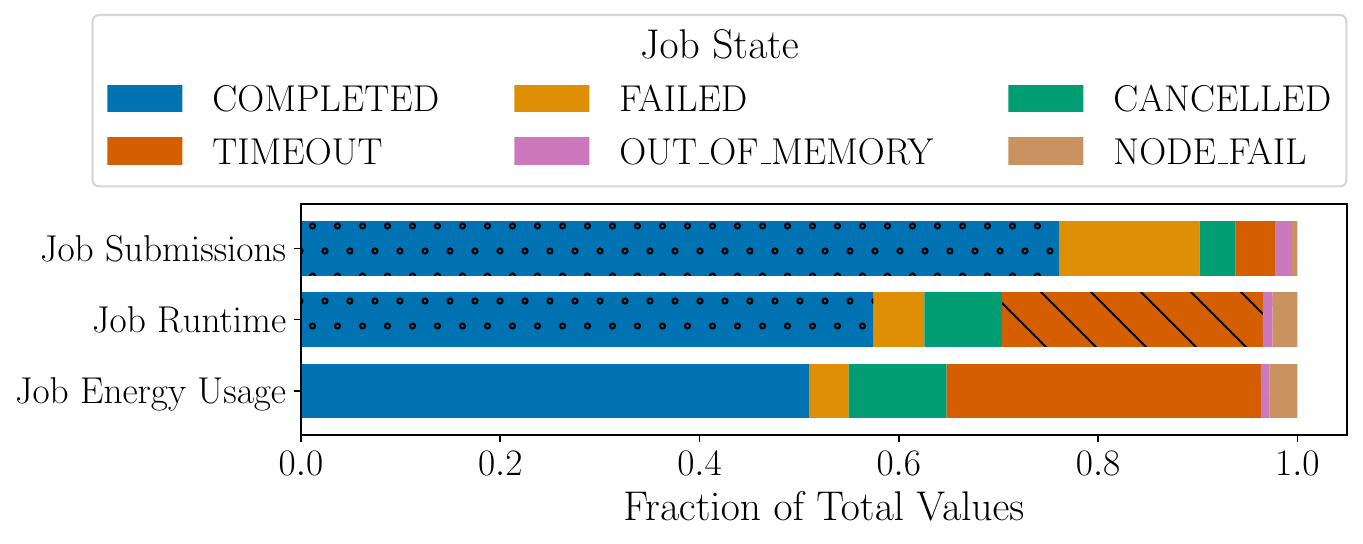}
    \caption{Generic Jobs.}
    \label{fig:job-state-fractions-generic}
  \end{subfigure}
  
  \begin{subfigure}[b]{\columnwidth} %
    \includegraphics[width=\textwidth]{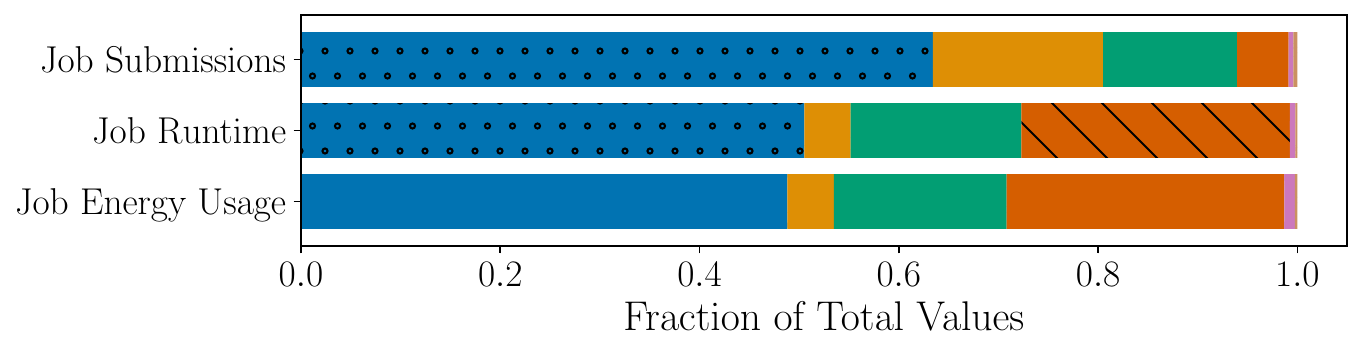}
    \caption{ML Jobs.}
    \label{fig:job-state-fractions-ml}
  \end{subfigure}
  \vspace*{-0.35cm}
  \caption{Fraction of job states, generic and ML jobs. About half the cluster-wide energy is spent on uncompleted jobs.}
  \label{fig:job-state-characterization}
  \vspace*{0.25cm}
\end{figure}

The combined dataset (Table\ref{table:data-preparation}-(c)) enables diverse cross-metric analyses between job and node traces, including energy consumption and correlation analysis.

\vspace*{-0.25cm}
\begin{tcolorbox}[enhanced, 
    breakable]
\defmainfinding{mf:cross-metric}
{Unsuccessful jobs consume about half of the total cluster energy. 
Concurrent jobs on the same node show correlations for terminating in the same state, especially generic jobs.}
{Checkpointing~\cite{DBLP:conf/dsn/GargPCT18} should be used to save partial work to avoid energy wastage due to failures.
Understanding job exit state correlations can enhance failure prediction mechanisms.}

\end{tcolorbox}

\subsection{Job Submissions, Runtime, and Energy Usage}\label{sec:cross:job-states-and-power-usage}

\begin{table}[t]
\centering
\caption{Job metrics distribution cluster-wide.}
\label{tab:ml_generic_percentages}
\vspace*{-0.2cm}
\begin{tabular}{llll}
\toprule
 Type    & Job Submissions   & Job Runtime   & Job Energy   \\
\hline
 Generic & 90.72\%            & 83.18\%        & 61.32\%       \\
 ML      & 9.28\%             & 16.82\%        & 38.68\%       \\
\bottomrule
\end{tabular}
\end{table}

\vspace*{0.06cm}
\noindent\fbox{%
    \parbox{\linewidth}{%
    \defobservation{ob:}
    {ML jobs have a smaller share of job submissions (9\%) and runtime (17\%) on the cluster but contribute to a relatively larger total energy footprint (39\%).}
    }}
\vspace*{0.03cm}

Table~\ref{tab:ml_generic_percentages} provides a cluster-wide overview of the distribution of generic and ML jobs.
Despite generic jobs accounting for 91\% of the total submissions and 83\% of the cumulative runtime, ML jobs demand a disproportionately higher amount of energy, consuming 39\% of the cluster-wide energy budget, while generic jobs account for 61\%.

Examining the distribution of job termination states within each type of job in Figure~\ref{fig:job-state-characterization} reveals some more details. Completed jobs make up the majority of submitted jobs, with a fraction of 76\% in Figure~\ref{fig:job-state-fractions-generic} for generic ones, and 63\% in Figure~\ref{fig:job-state-fractions-ml} for ML types. The fraction of submitted jobs failing is smaller for the generic type with 14\% than ML with 17\%. The biggest difference can be found in the fractions of canceled jobs, which are roughly 4\% and 13\% for submitted generic and ML jobs, respectively. The overall distribution of submitted generic jobs' exit states aligns with the findings of~\cite{DBLP:conf/sc/AnticiABK23} within a few percentage points. However, they diverge noticeably when comparing them to our spread of ML job states, further emphasizing the importance of looking at ML workloads separately.

Moving on to the sum of job runtimes, job states' proportions shift noticeably. Runtimes for jobs ending in a timeout state take up significant fractions of around 26-27\% for both generic and ML types, at the cost of smaller fractions for completed and failed jobs, with similar results being evidenced in the work of~\cite{DBLP:journals/fgcs/VersluisCGLPCUI23}. The trend of ML jobs having a relatively higher proportion of canceled jobs seen for job submissions continues for job runtimes. Interestingly, the share of runtime used for jobs exiting with the node failure state is 2.5\% among generic jobs, over 10 times higher than the 0.2\% among ML jobs.

\vspace*{0.06cm}
\noindent\fbox{%
    \parbox{\linewidth}{%
    \defobservation{ob:}
    {About 50\% of the total cluster energy is used for jobs terminating unsuccessfully.}
    }
}
\vspace*{0.03cm}

Cumulative energy usage of jobs grouped by state shows very similar patterns to their runtimes, with the fraction of timed-out jobs growing even larger. One key insight from Figure~\ref{fig:job-state-characterization} is that even though the majority of submitted jobs are completed successfully, about half of the used energy is spent on jobs resulting in unsuccessful terminations, like failures, timeouts, out-of-memory, or node failures. This highlights the huge potential for energy savings, e.g., by analyzing jobs more intensively before submission or implementing early-stopping mechanisms for long-running jobs that end up in a timeout.

\subsection{Correlation of Job States}\label{sec:cross:corr-job-states}
\vspace*{0.06cm}
\noindent\fbox{%
    \parbox{\linewidth}{%
    \defobservation{ob:}
    {High correlations exist between identical job states, indicating that jobs running concurrently on the same node tend to end in the same states. The highest correlations exist for the 'NODE\_FAIL' state, with values of 0.94 and 0.75, for generic and ML jobs, respectively.}
    }}
\vspace*{0.03cm}

\begin{figure*}[t] %
  \centering
  \begin{subfigure}[b]{0.49\textwidth}
    \includegraphics[width=\textwidth]{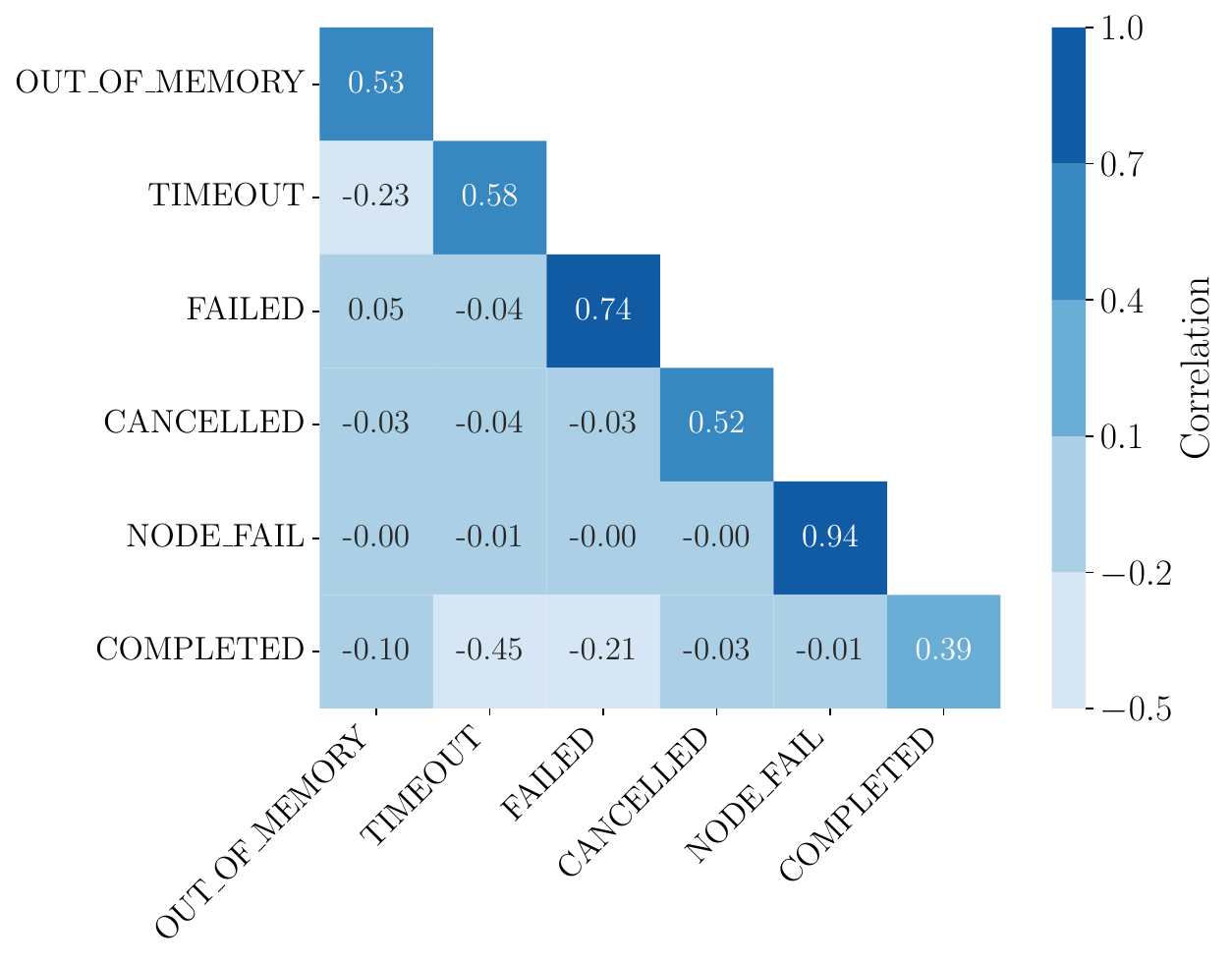}
    \vspace{-0.7cm}
    \caption{Generic Jobs.}
    \label{fig:job-state-correlation-generic}
  \end{subfigure}
  \hfill
  \begin{subfigure}[b]{0.49\textwidth}
    \includegraphics[width=\textwidth]{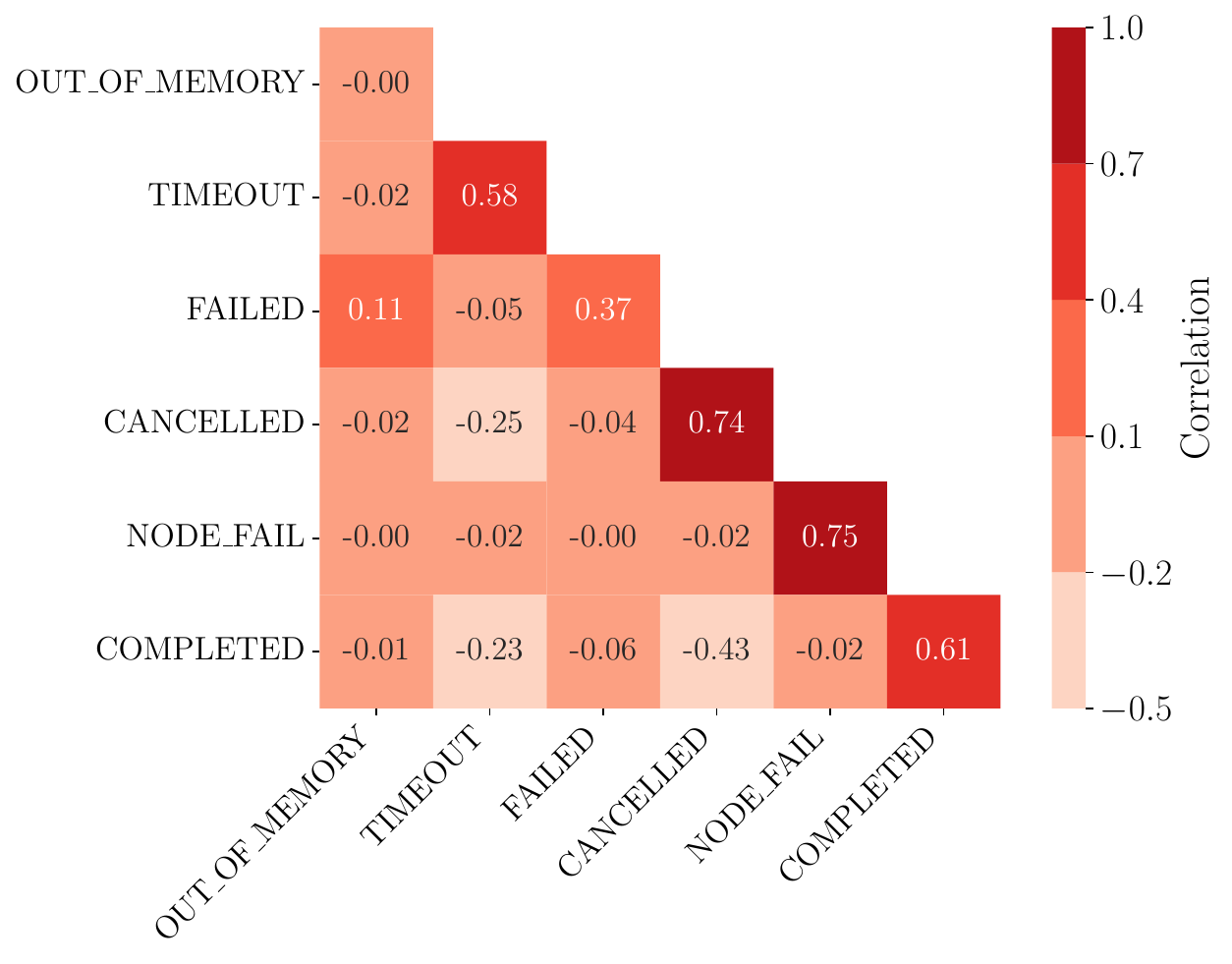}
    \vspace{-0.7cm}
    \caption{ML Jobs.}
    \label{fig:job-state-correlation-ml}
  \end{subfigure}
  \vspace*{-0.15cm}
  \caption{Job exit state correlation under high node load. Concurrent jobs tend to terminate in the same exit state.}
  \label{fig:job-state-correlation}
  \vspace*{-0.15cm}
\end{figure*}

We investigate whether concurrent jobs fail simultaneously by utilizing the combined dataset to correlate the exit states of two concurrently running jobs on the same node at the same timestamp. This novel analysis checks whether the job exit state correlates with that of other concurrent jobs using Pearson's correlation.
Naive usage of Pearson correlation can result in a spurious high correlation, as all the low usage periods could correlate with low job periods. To avoid this, we first find all periods with high power load using peak detection~\cite{yigitbasi2010analysis}, by filtering the dataset for node power usages one standard deviation above the mean.
We then correlate the termination states of concurrent jobs across all timestamps.
\par \Cref{fig:job-state-correlation} gives heatmaps for correlations between different job states, split up by generic and ML jobs. 
In general, there are high correlation coefficients between the same job states, as illustrated on the diagonals of the plots, meaning concurrent jobs are more likely to end up in the same state rather than different ones.
However, this observation does not always hold. While generic jobs do have a high correlation of 0.53 for the 'OUT\_OF\_MEMORY' state in~\Cref{fig:job-state-correlation-generic}, ML jobs in \Cref{fig:job-state-correlation-ml} show no correlation at all. One potential reason for this divergence can be the higher memory demand of generic jobs, compared to ML jobs. 
Job failures are also more highly correlated among generic jobs than ML jobs, with a correlation of 0.74 compared to 0.37.
Correlations are commonly high for generic and ML jobs for the states 'TIMEOUT' and 'CANCELLED',  with correlation values above 0.5. 'COMPLETED' jobs show a higher correlation among ML jobs than generic ones.  The strongest correlations between different states are negative with a value of -0.45 between the states 'COMPLETED' and 'TIMEOUT' for generic jobs, and a value of -0.43 between 'COMPLETED' and 'CANCELLED' for ML jobs.

One potential reason for the observed correlations could be that concurrent jobs are submitted by the same users. However, the job dataset lacks user information due to privacy constraints, limiting further investigation of this hypothesis.

\section{Scope and Threats to Validity}\label{sec8:threats}

Although this study offers new insights into the characterization of generic and ML workloads, we acknowledge that our findings are limited to the specific system and data we used.
The threats to the validity include: (1) \textit{Generalizability}: 
This study characterizes traces from only one datacenter. This is broadly taken in other research works (see \Cref{tab:related_work})
because of limited access
to publicly available data with similar granularities and timelines. 
Therefore, we use different strategies such as providing a guide to the characterization method, 
and comparing similar or different findings in other research works. These are commonly used techniques in literature ~\cite{DBLP:journals/fgcs/VersluisCGLPCUI23,DBLP:conf/usenix/AmvrosiadisPGGB18,DBLP:conf/sc/ShinOKEW21,li2022ai}. 
(2) \textit{Causality}: It is unlikely this kind of study alone can uncover causal relationships between the variables found to be correlated, but it provides corroborating evidence that such correlations exist and thus enables future root-cause analysis. 
(3) \textit{Categorization}: In our work, %
we are limited by the knowledge about the application domains of our ML workloads, hence, we look at ML jobs as a single category.
While this categorization is sufficient for many types of characterization, it potentially hides other detailed findings. Consequently, dividing ML jobs into multiple distinct sub-classes can help to mitigate this issue, such as grouping ML jobs by science domains~\cite{DBLP:conf/mascots/PaulKW21}, types~\cite{weng2022mlaas}, and maturity~\cite{li2022ai}.

\section{Related Work}\label{sec8:related-work}

\begin{table*}[]
    \caption{Comparison of related works. Legend: \#DC=The number of datacenter systems, \#U=Utilization analysis, \#E=Energy analysis, \#F=Failures analysis, \#J=Joint analysis on the relation between node and job data.}
    \label{tab:related_work}
    \vspace*{-0.2cm}
\normalsize
\renewcommand{\arraystretch}{1.3}
\resizebox{\textwidth}{!}{
\begin{tabular}{cllccccccccc}
\toprule
\multirow{2}{*}{\textbf{Year}} & \multirow{2}{*}{\textbf{Work}}                             & \multirow{2}{*}{\textbf{Characteristics}}                                       & \multicolumn{3}{c}{\textbf{Data Scope}}                                            & \multicolumn{2}{c}{\textbf{Job Types}}                             & \multicolumn{4}{c}{\textbf{Characterization Types}}                                                                                                                  \\ \cline{4-12} 
                              &                                                            &                                                                                 & \multicolumn{1}{c}{\textbf{\#DC}} & \multicolumn{1}{c}{\textbf{Job}} & \multicolumn{1}{c}{\textbf{Node}}  & \multicolumn{1}{c}{\textbf{Generic}} & \multicolumn{1}{c}{\textbf{ML}} & \multicolumn{1}{c}{\textbf{\#U}} & \multicolumn{1}{c}{\textbf{\#E}} & \multicolumn{1}{c}{\textbf{\#F}} & \multicolumn{1}{c}{\textbf{\#J}} \\ \hline
2013                                               & \cite{DBLP:conf/dsn/El-SayedS13}             & Characterizing node failures and related factors on a large-scale HPC.                                                                               & 1                                          & \cmark                            & \cmark                                  & \cmark                                    & \xmark                               & \cmark                                        & \cmark                                    & \cmark                                      & \cmark                                           \\
2014                                               & \cite{DBLP:conf/cluster/VermaKW14}                & Evaluating job packing in four different metrics.                                                                               & 1                                          & \cmark                              & \xmark                                   & \cmark                                    & \xmark                                & \cmark                                        & \xmark                                   & \xmark                                     & \xmark                                           \\
2017                                               & \cite{DBLP:conf/sc/GuptaPET17}            & Comparing failure characteristics of multiple largescale HPC systems.                                                                               & 5                                          & \xmark                      & \cmark                                & \cmark                                    & \xmark                               &  \xmark                                       & \xmark                                   & \cmark                                     & \xmark                                           \\
2017                                               & \cite{DBLP:conf/sosp/CortezBMRFB17}                 & Characterization and prediction of cloud VM workloads.                                                                               & 1                                          & \cmark                         & \xmark                               & \cmark                                    & \xmark                               & \cmark                                        & \xmark                                   & \cmark                                     & \xmark                                           \\
2018                                               & \cite{DBLP:journals/jpdc/RodrigoOEAGR18}                 & HPC workload characterization focus on job geometry and groups.                                                                               & 3                                          & \cmark                    & \xmark                                & \cmark                                    & \xmark                               & \cmark                                        & \xmark                                   & \xmark                                     & \xmark                                           \\
2018                                               & \cite{DBLP:conf/usenix/AmvrosiadisPGGB18} & Characterization of diverse cluster workloads and its impact on research.                                                                               & 4                                          & \cmark                                  & \xmark                               & \cmark                                    & \xmark                               & \cmark                                        & \xmark                                   & \cmark                                     & \xmark                                           \\
2019                                               & \cite{DBLP:conf/iwqos/GuoCWDFMB19}                      & Resource efficiency limitation analysis of Alibaba datacenter traces.                                                                               & 1                                          & \cmark                              & \cmark                                   & \cmark                                     & \xmark                                & \cmark                                         & \xmark                                    & \xmark                                      & \cmark                                            \\
2020                                               & \cite{DBLP:conf/ics/LiuLKHCRFP20}            & HPC job characterization/identification at leadership computing facility.                                                                               & 1                                          & \cmark                              & \cmark                                & \cmark                                    & \xmark                               & \cmark                                        & \xmark                                   & \cmark                                     & \cmark                                           \\
2020                                               & \cite{DBLP:conf/sc/PatelLKRAT20}                 & Long-term analysis of job characteristics on large-scale systems.                                                                               & 2                                          & \cmark                         & \xmark                             & \cmark                                     & \xmark                                & \cmark                                         & \xmark                                    & \xmark                                      & \xmark                                            \\
2020                                               & \cite{power_jobs_ipdps_20}             & Power consumption behavior analysis of jobs on HPC clusters.                                                                               & 2                                          & \cmark                        & \cmark                              & \cmark                                    & \xmark                               & \cmark                                        & \cmark                                   & \xmark                                     & \cmark                                           \\
2020                                               & \cite{DBLP:journals/tpds/IlagerRB21}                  & Thermal prediction for efficient energy management of clouds using ML.                                                                               & 1                                          & \xmark                             & \cmark                                 & \cmark                                    & \xmark                               & \cmark                                        & \cmark                                   & \xmark                                     & \xmark                                           \\
2021                                               & \cite{DBLP:conf/sc/ShinOKEW21}            &
 Power/energy/thermal analysis of a 200PF pre-exascale supercomputer.                                                                        & 1                                          & \cmark                      & \cmark                                & \cmark                                    & \xmark                               & \cmark                                        & \cmark                                   & \cmark                                     & \cmark                                           \\
2021                                               & \cite{DBLP:conf/mascots/PaulKW21}             & Characterizing machine learning I/O workloads on leadership-scale HPC.                                                                               & 1                                          & \cmark                                 & \xmark                                  & \cmark                                    & \cmark                               & \xmark                                        & \xmark                                   & \xmark                                     & \xmark                                           \\
2022                                               & \cite{li2022ai}                           & AI-workflow classification and analysis on GPU-accelerated systems.             & 1                                          & \cmark                              & \cmark                                 & \cmark                                    & \cmark                               & \cmark                                        & \cmark                                   & \xmark                                     & \cmark                                           \\
2022                                               & \cite{weng2022mlaas}                      & MLaaS characterization on a large-scale heterogeneous GPU-cluster.              & 1                                          & \cmark                                   & \cmark                                   & \xmark                                    & \cmark                               & \cmark                                        & \xmark                                   & \xmark                                     & \xmark                                           \\
2023                                               & \cite{DBLP:conf/wosp/ChuTVI23}                          & Job failure analysis of datacenter with mixed generic/ML workload.              & 1                                          & \cmark                           & \xmark                                 & \cmark                                    & \cmark                               & \xmark                                        & \xmark                                   & \cmark                                     & \xmark                                           \\
2023                                               & \cite{DBLP:journals/fgcs/VersluisCGLPCUI23}                   & Holistic characterization of both generic and ML job and node data.             & 1                                          & \cmark                      & \cmark                              & \cmark                                    & \cmark                               & \cmark                                        & \cmark                                   & \cmark                                     & \xmark                                           \\
2023                                               & \cite{ilager_dc_analysis_UCC23}        & Statistical driven datacenter workload analysis of energy and temperature.     & 1                                          & \xmark                              & \cmark                                & \cmark                                    & \xmark                               & \cmark                                        & \cmark                                   & \xmark                                     & \xmark                                           \\
2023                                               & \cite{DBLP:conf/sc/AnticiABK23}           & Large-scale HPC job power consumption dataset construction and analysis.                                                                               & 1                                         & \cmark                       & \cmark                         & \cmark                                    & \cmark                               & \xmark                                        & \cmark                                   & \cmark                                     & \cmark         \\                                  
2023                                               & \cite{li2023_nersc/10.1007/978-3-031-32041-5_16}           & Analyzing resource utilization in a heterogeneous large-scale HPC system.                                                                               & 1                                         & \cmark                       & \cmark                         & \cmark                                    & \cmark                               & \cmark                                        & \xmark                                   & \xmark                                     & \cmark                                           \\ \midrule
\multicolumn{1}{l}{\textbf{2024}}                  & \textbf{\textbf{Our work }}               & \textbf{Generic/ML workloads, utilization, energy, failures, and joint analysis.} & \multicolumn{1}{c}{\textbf{1}}             & \textbf{\cmark}                               & \textbf{\cmark}                        & \textbf{\cmark}                           & \textbf{\cmark}                      & \textbf{\cmark}                               & \textbf{\cmark}                          & \textbf{\cmark}                            & \textbf{\cmark}                                  \\ \bottomrule
\end{tabular}}
\vspace*{-0.25cm}
\end{table*}

We briefly summarize comparable related work in Table~\ref{tab:related_work}. %
Relatively, ours is the first to intensively compare generic and ML workloads, covering broadly analyzed metrics, such as utilization, energy, and failures, to the less commonly studied joint characterization of hardware and workload traces.

Most HPC datacenter studies primarily focus on generic workloads~\cite{DBLP:conf/dsn/El-SayedS13,DBLP:conf/cluster/VermaKW14,DBLP:conf/sc/GuptaPET17,DBLP:conf/sosp/CortezBMRFB17,DBLP:journals/jpdc/RodrigoOEAGR18,DBLP:conf/usenix/AmvrosiadisPGGB18,DBLP:conf/iwqos/GuoCWDFMB19,DBLP:conf/ics/LiuLKHCRFP20,DBLP:conf/sc/PatelLKRAT20,power_jobs_ipdps_20,DBLP:journals/tpds/IlagerRB21,DBLP:conf/sc/ShinOKEW21,ilager_dc_analysis_UCC23}, without splitting either jobs types (generic vs ML) or node types (CPU-only vs GPU nodes), as we do in our study.
Moreover, we also carry out joint analysis on the combined job and node data, while most other works do not consider this interplay, investigating job and node data independently~\cite{DBLP:conf/cluster/VermaKW14,DBLP:conf/sc/GuptaPET17,DBLP:conf/sosp/CortezBMRFB17,DBLP:journals/jpdc/RodrigoOEAGR18,DBLP:conf/usenix/AmvrosiadisPGGB18,DBLP:conf/sc/PatelLKRAT20,DBLP:journals/tpds/IlagerRB21,DBLP:conf/mascots/PaulKW21,weng2022mlaas,DBLP:conf/wosp/ChuTVI23,DBLP:journals/fgcs/VersluisCGLPCUI23,ilager_dc_analysis_UCC23}. Most of the  studies, focus their analysis on a single datacenter (see Table~\ref{tab:related_work}). 
Works that do analyze multiple datacenters together mainly focus on hardware utilization~\cite{DBLP:journals/jpdc/RodrigoOEAGR18,DBLP:conf/usenix/AmvrosiadisPGGB18,DBLP:conf/sc/PatelLKRAT20,power_jobs_ipdps_20}, 
or evaluate hardware and workload traces separately~\cite{DBLP:conf/sc/GuptaPET17,DBLP:journals/jpdc/RodrigoOEAGR18,DBLP:conf/usenix/AmvrosiadisPGGB18,DBLP:conf/sc/PatelLKRAT20}, 
and commonly do not holistically compare generic and ML based workloads~\cite{DBLP:conf/sc/GuptaPET17,DBLP:journals/jpdc/RodrigoOEAGR18,DBLP:conf/usenix/AmvrosiadisPGGB18,DBLP:conf/sc/PatelLKRAT20,power_jobs_ipdps_20}.

Shin et al. \cite{DBLP:conf/sc/ShinOKEW21} discuss utilization, the impact of GPU placement on power/temperatures, failure characterization, and cross-analysis of job and node data. However, they do not distinguish between generic and ML workloads like we do in our work.
Similar to our work, \cite{DBLP:conf/sc/AnticiABK23} conducts various workload characterizations, concerning job exit status, time, and energy, while also categorizing jobs into the CPU-only and CPU+GPU classes. Still, they do not show overall hardware utilization, which we add to complement our workload analysis.
Previous analysis of our datacenter~\cite{DBLP:journals/fgcs/VersluisCGLPCUI23} also looked at generic and ML workloads, but not in a joint job-node fashion as we did.
Li et al.~\cite{li2022ai} focus more on characterizing ML workloads and GPU-accelerated hardware, with less emphasis on comparison with generic workloads or job failure analysis.
The work of~\cite{li2023_nersc/10.1007/978-3-031-32041-5_16} contrasts GPU and CPU workloads but skips energy and failure analyses, which we include in our study.

\section{Conclusion}\label{sec9:conclusion}

In this work, we identified the emerging challenge of understanding ML workloads in contrast to general HPC workloads. 
We collected and released job-level and node-level data from a relevant HPC datacenter. Integrating job- and node-data sources, we analyzed utilization, energy, and failure occurrence, and also conducted a joint analysis to reveal the relation between job and node metrics. Our statistical characterization led to 11 major observations, contributing to 3 major findings and 3 actionable insights. %
Our findings help understand the impact of ever-growing ML jobs on HPC datacenters and provide valuable insights for datacenter operators. 
We released our datasets and software as open-access artifacts to encourage further research. %

\section*{Acknowledgment}
We thank the Dutch National Supercomputing Center SURF for
providing the data. We thank the China Scholarship Council (CSC)
for supporting Xiaoyu Chu. We thank the support of Netherlands-funded projects NWO OffSense and GFP 6G FNS, and EU-funded projects MCSA-RISE Cloudstars and Horizon Graph-Massivizer. This research has been partially funded through the projects: High-Performance Integrated Quantum Computing (HPQC), Austrian Research Promotion Agency (FFG) \# 897481; Transprecise Edge Computing (Triton), Austrian Science Fund (FWF), DOI: 10.55776/P36870; and Trustworthy and Sustainable Code Offloading (Themis), Austrian Science Fund (FWF), DOI: 10.55776/PAT1668223. %

\bibliographystyle{ieeetr}
\bibliography{reference}

\end{document}